\documentstyle[12pt]{article}

\def\br{\begin{thebibliography}{abc}}
\def\er{\end{thebibliography}}

\begin{document}

\thispagestyle{empty}
\setcounter{page}{0}
\bibliographystyle{unsrt}
\footskip 1.0cm
\thispagestyle{empty}
\setcounter{page}{0}
\begin{flushright}
SU-4240-599\\
December, 1994 \\
hep-th/9412220
\end{flushright}
\vspace{10mm}

\centerline {\bf NONCOMMUTATIVE GEOMETRY AND A DISCRETIZED VERSION }
\centerline {\bf  OF KALUZA-KLEIN THEORY WITH A FINITE FIELD CONTENT}
\vspace*{15mm}
\centerline {\bf Nguyen Ai Viet and Kameshwar C.Wali }
\vspace*{5mm}
\centerline {\it  Department of Physics, Syracuse University,}
\centerline {\it Syracuse, NY 13244-1130, U.S.A.}
\vspace*{15mm}
\normalsize
\centerline {\bf Abstract}
We consider a four-dimensional space-time supplemented by two discrete points
assigned to a
$Z_2$ algebraic structure and develop the formalism of noncommutative geometry.
By setting up a generalised vielbein, we study the metric structure. Metric
compatible torsion free connection defines a unique finite field content in the
model and leads to a discretized version of Kaluza-Klein theory. We study some
special cases of this model that illustrate the rich and complex structure
with massive modes and the possible presence of a cosmological constant.

\bigskip

Pacs numbers:  04.20.Jb, 04.40. +c, 11.15. -q, 14.80.Hv
\newpage

\section{ Introduction}
Mathematical framework based on classical differential manifolds and the
associated
algebras of smooth functions and their differentiable structures has provided
so far the necessary algebraic and geometric tools to construct quantum field
theories to describe elementary particle interactions. However, inspite of
great progress in our understanding of these interactions, problems remain. The
inadequacies of the Standard Model as a fundamental theory are too well
emphasized in the literature to merit repetition here. String theory's claim as
a fundamental theory that unifies all interactions including gravity is still
far from being established. It certainly has not made a convincing contact with
the experimental world and has not provided so far new insights into the
successes of the Standard Model. These and other considerations
beg for new mathematical ideas and new ways to explore physics at small scales.

Connes' recent developement of noncommutative geometry ( NCG)\cite{Co1,Co2}
has provided such new ideas and new tools to construct particle physics models
based on a geometric picture in which Higgs fields can be introduced as
geometric objects on an equal footing with the gauge fields. The Higgs fields
trigger spontaneous symmetry breaking in a natural way in the Standard Model
and Grand Unified Theories (GUT's)\cite{CoLo,ChamFF1,BGW}. Among several
approaches
to NCG, Connes' approach that we will follow in this paper is based on
enlarging the usual four-dimensional space-time by including additional
discrete dimensions. This leads to more than one copies of space-time and
enables one to introduce different symmetries on different copies.

Connes' NCG can also be thought of as describing a discretized version of
Kaluza-Klein theories \cite{KK} that ordinarily aim at incorporating internal
symmetries of elementary particles and unify their interactions. Instead of
compactified continuous space degrees of freedom, in NCG we have a countable
number of discrete points.  With this in mind, one may ask how to introduce
gravity in this space-time. The first step in this direction was taken by
Chamseddine, Felder and Fr\"ohlich \cite{ChamFF2}, who gave a generalization
of the basic notions - vierbein, spin connection and curvature - of Riemannian
geometry in the
context of the new framework. As a result, they obtained Einstein's gravity
along with Brans-Dicke scalar field. More recently, several others
\cite{Kast,KaWa,SI,China} have introduced some refinements, but have obtained
essentially the same results.

In this and a previous short paper \cite{LVW}, we have followed
a different track. Guided by the Kaluza-Klein theory, we consider the two-point
internal space as a discretization of the fifth dimension.
We keep all the allowed fields by assuming the most general form for
the vielbein and therefore, vector and scalar fields appear naturally along
with tensor fields \footnote{ Klim\~cik et al \cite{Klim}, independently, have
considered gravity together with a vector field ( without the Brans-Dicke
field). Madore \cite{Madore} has also discussed the possibility of
having vectors fields together with gravity in a somewhat different NCG
with matrix algebras $M_n$}. With two discrete point internal space, these
fields come in pairs and in each pair, one field is massless and the other is
massive. In the
conventional Kaluza-Klein theories, the particle spectrum consists of infinite
number of massive modes. Usually, one resorts to truncation of the spectrum,
but then one runs into inconsistencies in constructing
realistic theories \cite{PoDu}.

In the previous paper \cite{LVW}, we had a restricted version of the theory
due to the choice of a hermitian connection. We showed that
the zero mode sector of the Kaluza-Klein theory emerged without truncation. In
this extended version, we consider a general formalism dictated by NCG. We
shall
see that the torsion free and metric compatibility conditions impose a very
strong constraint on the field content of the theory. In the more general case,
these conditions allow, beside the tensor, vector and scalar fields, two new
dynamical fields $\alpha(x)$ and $ \beta(x)$. The $\beta(x)$ field rescales the
metric on one sheet and therefore it acts like a dilaton field in conventional
theories. However, we find it has a mass term unlike in the conventional
theories. Furthermore, if we assumed
$\beta $ to be a constant everywhere, it gives rise to a
cosmological constant. Similarly, the $\alpha(x)$ field rescales the vector and
the scalar fields  on one sheet and when we assume it to be constant, it makes
the vector field massive.
{}From physical considerations, if we adhere to metric compatibility, then
torsion is not arbitrary, but it is determined in terms of the metric and
allows to have both massless and massive fields. In other words, a metric
compatible torsion provides the {\it raison d'\^etre} for the massive modes in
a theory that has originally only zero modes. This is a new and beautiful
feature of NCG.

The paper is organized as follows: In the next section, we extend the
formalism in \cite{LVW} by including more details about the two-point internal
space and noncommutative differential calculus. In  Section 3, we set up
the vielbein in an orthonormal basis and discuss the structure of the metric
that follows. We also give the definitions of various inner products that are
necessary in our computations. In Section 4, after defining the generalized
connection, covariant derivative and torsion, we use the metric compatibility
condition to obtain the torsion free connection which can be used to compute
the generalised Ricci scalar curvature and the Lagrangian. Section 5. is
devoted
to some special cases and the final section to a summary and discussion of the
results.

\setcounter{equation}{0}
\section{ Two-point internal space, noncommutative differential calculus}
\subsection{ Basic elements of noncommutative geometry}
Let us consider a physical space-time manifold
${\cal M}$ extended by a discrete internal space of two points to which we
assign a $Z_2$-algebraic structure. Hence, besides the space-time variable, we
will have a new discrete variable denoted by an element
$ h \in Z_2 = \{ e,r~|~ e^2=e~,~ r^2=e~,~ er=re=r ~\}$.
With this extended space-time, the customary algebra of smooth functions
 ${\cal C}^\infty ({\cal M})$ is generalized to ${\cal A} ~=~
{\cal C}^\infty({\cal M})\oplus{\cal C}^\infty({\cal M})$ and any
generalized function $F \in {\cal A}$ can be written as
\begin{equation}
F(x)= f_+(x) e + f_-(x) r~ ,
\end{equation}
which can be viewed as a formal expansion by the $ Z_2$ variable.
We can represent the elements of the $ Z_2$ algebra by $ 2 \times 2$ matrices:
\begin{equation}
e =\pmatrix{1&0\cr
                0&1\cr}  ~~ ,~~ r=\pmatrix{1&0\cr
                                                 0&-1\cr}~.
\end{equation}
Then the function $ F(x)$ is represented by a $2 \times 2$ matrix
\begin{equation}
F  ~=~f_+(x)\pmatrix{1&0\cr
       0&1\cr} ~ + ~ f_-(x) \pmatrix{1&0\cr
                                   0&-1\cr} ~=~\pmatrix{f_1(x)&0\cr
                                                0&f_2(x)\cr}~,
\end{equation}
where $f_1, f_2$ are obvious combinations of $ f_+, f_-$.
To simplify notation, we will use the same mathematical symbols for abtract
elements and their representations. In this paper we will use the small
letters to denote the quantities of ordinary geometry and capital letters
for generalized quantities of NCG.

The algebra ${\cal A}$ of smooth functions can be considered as the algebra
of the generalized 0-forms $ \Omega^0({\cal M})= {\cal C}^\infty({\cal M})
\oplus {\cal C}^\infty({\cal M})$. To go beyond the ordinary geometry, we
must introduce a second important geometric ingredient, the Dirac operator
$ D$ \cite{Co1,Co2}, that serves as an exterior derivative giving us the
starting point of the noncommutative differential calculus.
As a direct generalization of the usual exterior derivative $ d=
dx^\mu \partial_\mu~ ,~~d^2=0$, we will assume that it has the form
$ D = d + Q $, where $Q $ is the part of the exterior derivative, that comes
from derivations over the $Z_2$ internal variable. Connes has given
it \cite{Co2} formally as
\begin{equation}\label{DERDIS}
D:(f_1,f_2) \longrightarrow (df_1, df_2, m(f_2-f_1), m(f_1-f_2)),
\end{equation}
where m is a parameter with dimension of mass or the inverse of length.
Therefore, it is apparent that we can look upon the last two terms in Eq
(\ref{DERDIS}) as
representing derivatives over the discrete dimensions.

We are seeking a realization of Eq (\ref{DERDIS}) in which the operator
$ D$ appears more transparently as a derivative operator satisfying the
Newton-Leibnitz rule. For this purpose, let us define derivatives as
follows;
\begin{eqnarray}\label{CODER}
&& D_\mu = \pmatrix{\partial_\mu &0\cr
                          0 &\partial_\mu\cr} ,~~~\mu = 0,1,2,3~,
\cr
&& D_5 = \pmatrix{0& m\cr
-m&0\cr}  ,
\end{eqnarray}
and specify their action on the 0-form elements as
\begin{equation}\label{COMDER}
D_N(F)= [D_N, F] ~~,~~N=\mu, 5.
\end{equation}
It is easy to verify that in the above representation, $D_N$ satisfies the
Newton-Leibnitz rule,
\begin{equation}
D_N(FG) = D_N(F) G + F D_N(G).
\end{equation}
Hence, we can consider $D_N$ as derivations in the $Z_2$-noncommutative
geometry.

The generalized differential elements $DX^M$ have the following realizations
\footnote{ Here we have used a slightly different definition for $ DX^5$
compared with the one in Ref \cite{LVW}. This new choice for $DX^5$ is more
suitable to be a basis of 1-forms discussed in the next subsection.}
\begin{eqnarray}
DX^\mu & \doteq & \pmatrix{dx^\mu &0 \cr
                                     0& dx^\mu \cr} ,~~~ \mu = 0,1,2,3~,
 \nonumber\\
DX^5 \sigma^\dagger  & \doteq & \pmatrix{\theta& 0\cr
                                   0& -\theta \cr},
\end{eqnarray}
where $\theta $ is a Clifford element satisfying
\begin{equation}
\theta^2= 1~~~, ~~~ \theta dx^\mu = -dx^\mu \theta~.
\end{equation}
and
\begin{equation}
\sigma ~=~ \pmatrix{ ~0 & 1\cr
                    -1 & 0 \cr}
\end{equation}
The exterior derivative operator $D$ is given by
\begin{eqnarray}
D \doteq (~DX^\mu D_\mu~+~ DX^5 \sigma^\dagger D_5~)
\equiv \pmatrix{d&\theta m\cr
               \theta m & d\cr}~,
\end{eqnarray}
where $d$ denotes the exterior derivative on ${\cal M}$.
The exterior derivative acts on $F = (f_1, f_2)
\in \Omega^0({\cal M})$ as follows:
\begin{eqnarray}\label{DER0}
DF\doteq (~DX^\mu D_\mu ~+~DX^5 \sigma^\dagger D_5~)F =
                            \pmatrix{df_1& \theta m(f_2-f_1)\cr
                             \theta m(f_1-f_2) &df_2\cr} .
\end{eqnarray}

By placing the "discrete derivative" off-diagonal, we really mean that
$ Q $ is an outer automorphism. The importance of an outer automorphism was
discussed by Balakrishna et al in Ref.\cite{BGW}. Without this off-diagonal
part, even with an extended algebra of 0-forms the geometry has only the
commutative character as in the ordinary geometry \cite{KUMA}.

Working in the Hilbert space of spinors, Connes  \cite{Co2,CoLo}
choose the '$\Gamma$-representation' of differential elements
\begin{eqnarray}\label{dirac}
\Gamma^\mu = \pmatrix{\gamma^\mu & 0\cr
                        0& \gamma^\mu\cr} ,~~~
\Gamma^5 = \pmatrix{\gamma^5&0\cr
0&-\gamma^5\cr}.
\end{eqnarray}
To compare with papers that use the representation (\ref {dirac}), let us
note that our $ DX^\mu$ and $ DX^5\sigma^\dagger $ correspond to $\Gamma^\mu$
and $\Gamma^5$ respectively.
Hence, their Dirac operator in the $Z_2$-noncommutative geometry has the
self-adjoint realization
\begin{eqnarray}
\not \!\!D \doteq \Gamma^N D_N \equiv \pmatrix{\not\!\partial &\gamma^5
m\cr
                                           \gamma^5 m
&\not\!\partial\cr}~.
\end{eqnarray}
In our formalism $\theta $ is not necessarily $\gamma_5$ but can be
any Clifford element depending on the content of the matter field. As we are
working only with the pure geometric sector, we will not refer to its concrete
meaning. The triplet $({\cal A}, D, {\cal H})$, where ${\cal H}$ is a Hilbert
space of matter fields is the basic ingredient of NCG. We have already assumed
that ${\cal H}~=~{\cal H}_1 \oplus {\cal H}_2$ in representing the
algebra ${\cal A}$ and the Dirac operator $ D$ by $ 2 \times 2 $ matrices.
The algebra ${\cal A}$
completely replaces the concept of an underlying manifold. In NCG, calculations
can be done formally without explicit derivatives as in Eqs (\ref{CODER}) and
(\ref{COMDER}). We introduce such an object just to show that NCG is a
direct generalisation of the ordinary geometry, that was traditionally founded
on the notion of a tangent space at a point. It is worth noting that
Dubois-Violette, Kerner and Madore \cite{Madore} have also introduced
derivatives when discussing a different NCG with matrix algebras $M_n$.

\subsection{ Wedge product and generalized differential forms}
We can extend the space of derivatives of 0-forms in Eq (\ref{DER0}) to the
space of 1-forms  $\Omega^1({\cal M})$, where any 1-form $U\in
\Omega^1({\cal M})$ is defined as
\begin{eqnarray}
U\doteq DX^N U_N = \pmatrix{dx^\mu u_{1\mu}(x)& \theta u_2(x)\cr
                              \theta u_1(x) & dx^\mu u_{2\mu }\cr },
\end{eqnarray}
where $U^\mu, U_5$ are elements of $\Omega^0({\cal M})$.

Let us note that the hermitian conjungate of an 1-form is also an 1-form
\begin{equation}
U^{\dagger}= U_N^{\dagger}.DX^N = DX^\mu U_\mu + DX^5 {\tilde U_5},
\end{equation}
where we have introduced the notation $ {\tilde F} $:
\begin{equation}
 {\rm if}~~ F \in {\cal A}~~,~~ F=\pmatrix{f_1 & 0 \cr
              0 & f_2\cr} ~~,~~{\rm then}~ {\tilde F}= \pmatrix{ f_2 & 0 \cr
                                                     0 & f_1 \cr}.
\end{equation}

It is straightforward to generalize the definition of the wedge product to
construct
higher differential forms and differential algebra
by defining
\begin{eqnarray}\label{WEDGE}
DX^\mu \wedge DX^\nu & \doteq & \pmatrix{ dx^\mu \wedge dx^\nu & 0\cr
0& dx^\mu \wedge dx^\nu\cr}  \equiv  -DX^\nu \wedge DX^\mu ,\nonumber\\
DX^5 \wedge DX^\mu & \doteq & \pmatrix{ 0 & \theta dx^\mu \cr
  \theta dx^\mu & 0\cr} \equiv -DX^\mu \wedge DX^5 , \nonumber\\
DX^5 \wedge DX^5 &\doteq & 0
\end{eqnarray}
Alternately, we could have postulated $ DX^5 \wedge DX^5 \not= 0$ and
recover the construction of Coquereaux et al.\cite{COQUEV}. To do this, of
course, we have to assume that the additional dimension represented by the
discrete variables is an odd dimension to have a commuting differential
element instead of the anti-commuting one. Considering
our space-time as a discretized version of Kaluza-Klein theory,
we continue to treat the internal space as an even dimension on
an equal footing with the space-time coordinates and hence the wedge
product (\ref{WEDGE}).

A general $p$-form $W_p \in \Omega^p$ is defined as
\begin{equation}
W_p \doteq DX^{N_1}\wedge...\wedge DX^{N_p} W_{N_1 ... N_p}.
\end{equation}
where $ W_{N_1 ... N_p}$ are  generalised 0-forms.

The exterior derivative $DW_p \in \Omega^{p+1}$ of a $p$-form $W_p \in
\Omega^p$ is defined to be
\begin{equation}
DW_p ~ = ~ (DX^\mu \wedge DX^{N_1}\wedge....\wedge DX^{N_p} D_\mu + DX^5 \wedge
 DX^{N_1}\wedge ...\wedge DX^{N_p}D_5)W_{N_1...N_p},
\end{equation}
The wedge product $ W_{1p} \wedge W_{2q} \in
\Omega^{p+q}$ of a $p$-form $W_{1p}\in\Omega^p$ and a q-form
$W_{2q}\in \Omega^q$ is given as follows
\begin{equation}
W_{1p}\wedge W_{2q}  =  DX^{N_1}\wedge ...\wedge DX^{N_p}\wedge DX^{N_{p+1}}
\wedge...\wedge DX^{M_{p+q}}(W_1.W_2)_{ N_1...N_p N_{p+1}...N_{p+q}}.~~~
\end{equation}
where
\begin{equation}
(W_1.W_2)_{N_1...N_p N_{p+1}...N_{p+q}}~=~ W_{1 N_1...N_p}.
W_{2 N_{p+1}...N_{p+q}}~ \nonumber,
\end{equation}
if there is one $ 5 $ index among $ N_1,...,N_p $ indices, and
\begin{equation}
(W_1.W_2)_{N_1...N_p N_{p+1}...N_{p+q}}~=~\tilde W_{1 N_1...N_p}.
W_{2 N_{p+1}...N_{p+q}}~\nonumber ,
\end{equation}
if there is no $ 5 $ index among $ N_1,...,N_p $ indices.

We have the following essential properties for the exterior
derivative,
\begin{eqnarray}
&& D^2W_p = 0~,~~~~~\forall~ p~, \nonumber\\
&& D(W_p\wedge W_q) = DW_p \wedge W_q + (-1)^p W_p\wedge DW_q~.
\end{eqnarray}
The noncommutative character of our geometry is reflected in the fact that $W_p
\wedge W_q$ and $W_q\wedge W_p$ are not related in general to each other by a
simple factor as in the case of ordinary commutative geometry.

Although, in what follows, the geometrical objects we construct resemble
those of ordinary geometry, their noncommutative character dictates
a specific order in their definitions.
\subsection{ Inner product and signature in  "flat space-time"}
In 'flat' NCG, we can define the signature as
\begin{equation}
G^{MN}~=~ G(DX^M, DX^N)~=~< DX^M~,~DX^N>~=~ \eta^{MN},
\end{equation}
where $\eta ^{MN}=(-,++++)$. In the '$\Gamma $-representation'
the inner product is simply a Clifford trace ( not to be confused with the
trace over the $ 2 \times 2 $ matrix indices),
\begin{equation}
\eta^{MN} ~=~ {1 \over 4} Tr ({ \Gamma^M ~\Gamma^N})
\end{equation}
\setcounter{equation}{0}
\section{ The generalized metric and an orthonormal basis}
In this section we discuss the setting up of a generalized vielbein and the
ensuing metric structure of the assumed space-time.
\subsection{ The generalized vielbein}
In Riemann-Cartan geometry, the existence of a metric structure
on a manifold is equivalent to the assumption that there exists an orthonormal
basis of vierbein, that are 1-forms. We extend this idea to NCG and assume, as
in \cite{LVW} that there exists a generalized vielbein $\{E^A\}~ (A=a,\dot 5)$
. $ E^A \doteq DX^M E^A_M $  are 1-forms in the $Z_2$-noncommutative geometry
with the general form
\begin{eqnarray}
E^a &\doteq & \pmatrix{e^a_1& \theta f_2^a\cr
                   \theta f^a_1 & e_2^a\cr}~=~DX^\mu E^a_{~\mu}~+~DX^5 F^a
,~~ a = 0,1,2,3~,\nonumber\\
{}~~ \nonumber \\
E^{\dot 5} &\doteq & \pmatrix{a_1& \theta \phi_2\cr
                   \theta \phi_1 & a_2\cr}~=~DX^\mu A_\mu + DX^5 \Phi ,
\end{eqnarray}
where $e^a_1, e^a_2$ are vielbein on ${\cal M}$,
$a_1, a_2$ are 1-forms on ${\cal M}$ and
$f_1^a, f_2^a, \phi_1, \phi_2 $ are real functions on ${\cal M}$. We use a
$\dot 5$ index in the orthonormal basis to distinguish it from the index 5 in
the general one.

As in the usual Riemannian geometry, we still have a degree of freedom to
choose the following forms for vielbein without any loss of generality:
\begin{eqnarray}\label{gviel}
E^a &\doteq &\pmatrix{e^a_1& 0\cr
                   0 & e_2^a\cr} ~=~ DX^\mu E^a_\mu~~ a = 0,1,2,3, \nonumber\\
{}~~ \nonumber \\
E^{\dot 5} &\doteq & \pmatrix{a_1& \theta \phi_2\cr
                   \theta \phi_1 & a_2\cr}~=~DX^\mu A_\mu + DX^5 \Phi .
\end{eqnarray}

In Ref \cite{LVW} we have considered the self-adjoint vielbein
\begin{eqnarray}\label{svielb}
E^a &\doteq& \pmatrix{e^a& 0\cr
                   0 & e^a\cr} = DX^\mu e^a_\mu \nonumber\\
{}~~ \nonumber \\
E^{\dot 5} & \doteq & \pmatrix{a& \theta \phi \cr
                   \theta \phi & a\cr} = DX^\mu a_\mu + DX^5  \phi(x).
\end{eqnarray}
Here we will consider the general case (\ref{gviel}).

\subsection{ $DX^M$ basis and vielbein in two different representations}

In the last subsection $ DX^\mu$ and $DX^5$ have been chosen as pure diagonal
and pure off-diagonal respectively. Technically, it is convenient to choose
such a representation, in which the rules of differential calculus in the
previous section can still be used.
The basis $E^A$ can be used to formulate the structure equations and to
read the field content of the theory conveniently. However, in NCG
it is troublesome to use this basis to compute higher forms. The main reason is
that $E^A$ does not satisfy the anti-commutativity of the wedge
product for all its components, hence higher forms do not have a unique
expansion in this basis. Thus
\begin{eqnarray}
E^a\wedge E^b &~=~& - E^b\wedge E^a~, \cr
E^{\dot 5} \wedge E^{\dot 5} &~=~& E^b\wedge E^c A_bE^\mu_c a_{-\mu}r ~
-~ E^{\dot 5}\wedge E^cE^\mu_c a_{-\mu}r ~, \cr
E^{\dot 5}\wedge E^a &~=~& - E^b\wedge E^{\dot 5}{\tilde E}^\mu_b E^a_\mu ~+~
E^b\wedge E^c A_c({\tilde E}^\mu_b E^a_\mu - \delta ^a_b ) ~.
\end{eqnarray}

Spinors are defined locally in a locally flat basis of vielbein. Because of the
local flatness of the orthonormal basis, $E^a$ and $E^{\dot 5}$ can be
represented by 'flat' $\Gamma$ matrices as in Eq (\ref{dirac}). The trace
over the spinor indices can be taken only in this frame.

On the other hand, in a curved space the differential elements become
curvilinear.
In NCG, we will assume similarly that the basis of generalised 1-forms
$DX^\mu~,~DX^5$ will no longer be orthonormal in "curved" space time. In
general,
we can allow $DX^\mu$ and $DX^5$ to mix with each other. So the basis vectors
of the Hilbert space, in which $DX^M$ are pure diagonal or pure off-diagonal
will be combinations of two Hilbert spaces ${\cal H}_1$ and ${\cal H}_2$.
Hence, we cannot use the trace in the $DX^M$ basis but always have to go back
to the orthonormal basis to do so. In the basis $ DX^M$ the inner products
defined in the paragraph 3.4 should be used to compute the metric as we will
see later.

Finally, a comment about the existence of the vector $A_\mu$ is in order.
Intuitively, the existence of the vector field is related to the freedom to
assign to a pair of point on different
sheets the same coordinate system. The vector fields will appear if we mix
$DX^5$ and $DX^\mu$ to redefine $DX^5$ basis of the "curved" 1-forms as
\begin{equation}\label{gauge}
DX^5 \longrightarrow DX^5 - DX^\mu ~D_\mu \Lambda(x),
\end{equation}
where $\Lambda (x) $ is an arbitrary generalized function.
This degree of freedom will guarantee the gauge invariance of the generalised
interval defined in the next subsection.
\subsection{The Metric and its structure}
 Having a vielbein we can always construct a metric tensor $G$. We will think
of $G$ as a sesquilinear functional  \cite{Co2} $ G : \Omega^1 \times
\Omega^1 \longrightarrow {\cal A}~,$ having the hermitian structure
\begin{equation}\label{metric1}
G(U F, W H) = F^{\dagger}G(U, W) H ~, ~~~
\forall~~ U,W\in \Omega^1~,~ F, H \in \Omega^0.
\end{equation}

In the $E^A$-basis, the metric is taken to be
\begin{equation}\label{fmetric}
G(E^A, E^B) = \eta^{AB}~~,~~ \eta^{AB} = diag(-1,1,1,1,1)~.
\end{equation}
In the $DX^M$-basis we will have
\begin{equation}\label{cmetric}
G^{M N}=G(DX^M, DX^N) = E^{M~}_{~~A}\eta^{AB} E^N_{~~B}~,
\end{equation}
where $E^M_{~~A}$ are the inverses of  $E^A_{~~M}$.

 Explicitly the components of the metric are:
\begin{eqnarray}\label{KKMET2}
G^{\mu\nu} & = & \pmatrix{ g_1^{\mu\nu} & 0 \cr
                          0 & g_2^{\mu\nu} \cr} , \nonumber \\
G^{\mu 5} & = & - A^\mu\Phi^{-1}  , \nonumber \\
G^{5 \mu} & = &  - \Phi^{-1} A^\mu  , \nonumber \\
G^{55} & = & \Phi^{-2}(1+ A^2) .
\end{eqnarray}
where $g_i^{\mu \nu} = e^\mu_{ia}\eta^{ab} e^\nu_{ib}~,~ i=1,2$ are
the metrics on two sheets.
The inverse metric satisfying $G_{MN}.G^{NK}= G^{KN}.G_{NM}=\delta^K_{~N}$ is
\begin{equation}
G_{M N} = E_{~M}^{A}\eta_{AB} E_{N}^{B}~ .
\end{equation}
Explicitly,
\begin{eqnarray}\label{KKM2}
G_{\mu\nu} & = &  A_\mu A_\nu + \pmatrix{ g_{1\mu\nu} & 0 \cr
                                 0 & g_{2\mu\nu}\cr}   , \nonumber \\
G_{\mu 5} & = & A_\mu \Phi     , \nonumber \\
G_{5 \mu} & = & \Phi A_\mu       , \nonumber \\
G_{55} & = &  \Phi^2  .
\end{eqnarray}
It is worth noting that the metric continues to be symmetric, althought the
vielbein is not Hermitian.
\begin{eqnarray}
G_{MN} &~~=~~& G_{NM} , \nonumber \\
G^{MN} &~~=~~& G^{NM}.
\end{eqnarray}
Hence, we can define the generalised interval as
\begin{equation}
DS^2 ~~=~~ DX^M ~DX^N G_{MN}(x)
\end{equation}
It is easy to check that the generalised interval is invariant under the gauge
transformations (\ref {gauge}) and
\begin{equation}
A_\mu(x) ~ \longrightarrow ~~ A_\mu (x) + \partial_\mu\Lambda(x).
\end{equation}
This feature is exactly the same as in Kaluza-Klein theory.
In the locally flat frame, $E^A$ can be represented as flat $\Gamma$ matrices
in Eq (\ref{dirac}) and the metric can be computed by taking a trace over
the local spinor indices to give (\ref{fmetric}). This definition of metric
in the locally flat basis $ E^A $ is consistent with the metric defined in
the diagional represention of $DX^M$ via Eq (\ref{cmetric}).

\subsection{Inner products of forms and the volume element}

In our computations in the next section, we need to introduce inner product of
one- and two-forms and their extensions. For later convenience, we define them
here.
First, we will give the definitions of the inner products in the diagonal
representation of the $DX^M$ basis and then show that they are consistent with
the
definitions in the locally flat $\Gamma$ representation of the vielbein $E^A$.

To start with, we note that metric structure on a curved manifold defines an
inner product in the algebra
$\Omega^1({\cal M})$ of generalized 1-forms. We denote the metric as the
sesquilinear and hermitian inner product of two 1-forms,
\begin{equation}
 G^{MN}~=~ <~ DX^M~,~ DX^N~>.  \label{inner1}
\end{equation}
Then the inner product of two arbitrary 1-forms $U=DX^M U_M$ and $V=DX^N V_N$
can be computed from Eq (\ref{inner1}) as
\begin{equation}
< U, V> = U^\dagger_M<DX^M,DX^N> V_N= U^\dagger_N G^{MN}V_N.
\end{equation}

The first extension of the inner product (\ref{inner1}) we need is the inner
product of one
1-form and a tensorial product of two 1-forms. As in the case of the metric, we
will require that this innner product be sesquilinear and possesses a hermitian
structure. That is to say,
\begin{eqnarray}\label{OIN}
  <U\otimes V,~W>&~=~& V^{\dagger}~G(U~,~W),   \cr
  <U~,~W\otimes V> &~=~& G(U~,~W)~V,
\end{eqnarray}
where $U,V$ and $W$ are 1-forms.
Hence, this inner product in the $DX^M$ basis is
\begin{eqnarray}\label{INNER2}
<DX^M \otimes DX^N ,~ DX^P> &~=~& DX^N <DX^M,~DX^P>~=~DX^N~G^{MP},\cr
< DX^M ,~ DX^P\otimes DX^N > &~=~& <DX^M,~DX^P>DX^N~=~G^{MP}~DX^N .
\end{eqnarray}

The second extension of the inner product is for two 2-forms. It also has
a hermitian structure and is sequilinear, and again it is sufficient for our
purposes to
give it in the same basis as the first inner extension:
\begin{equation}\label {inner3}
     <U F~,~ W G > ~=~  F^{\dagger}~< U~,~W>~ G ,
\end{equation}
where $U,W$ are 2-forms and $ F,G$ are two arbitrary $ 2 \times 2$ matrices.
We shall define
\begin{equation}\label{INNER3}
<~DX^M\wedge DX^N~,~DX^R\wedge DX^S~> ~=~ {1\over 2} ( G^{MS}~G^{NR}~-~
G^{MR}~G^{NS}).
\end{equation}
and calculate (\ref{inner3}) by expanding the two-forms in the $DX^M$ basis.
All three inner products defined above are direct generalizations of
the corresponding products in the Riemannian geometry.

The above formulae are valid for calculations in the $DX^M$ basis.
In the orthonormal basis $E^A$, we can use another representation, where
$E^A$ are {\it not} in the representation (\ref{gviel}), but in the flat
$\Gamma$ representation (\ref{dirac}). In that representation, the inner
product can be taken as trace over the {\it local} spinor indices as dicussed
previously. This is consistent with our definition of inner product, since
\begin{equation}
G(E^A, E^B) ~=~ < E^A, E^B>,\nonumber
\end{equation}
and in the $\Gamma$-representation
\begin{equation}
G(E^A,E^B)~=~Tr(\Gamma^A \Gamma^B)=\eta^{AB},\nonumber
\end{equation}
In the representation (\ref{gviel})
\begin{equation}
G(E^A,E^B) ~=~ <DX^M E_M^A, DX^N E_N^B>= E^A_M G^{MN} E^B_N
\end{equation}
Due to Eq (\ref{cmetric}) the consistency is obvious.

The volume element is given by
\begin{equation}\label{volume}
D^5X~=~D^4X \sqrt{-det | G|}
\end{equation}
Here $det|G|$ denotes the determinant of our generalized metric defined in
Eq(\ref{volume}) and is given by
\begin{eqnarray}
det|G|& \doteq &{1\over 5!}{\epsilon }_{N_1 N_2 N_3 N_4 N_5}
{\epsilon}_{M_1 M_2 M_3 M_4 M_5} G^{N_1 M_1} G^{N_2 M_2} G^{N_3 M_3}
G^{N_4 M_4} G^{N_5 M_5} \nonumber\\
      & = &
{1\over 4!}{\epsilon}_{ \nu_1 \nu_2 \nu_3 \nu_4}
{\epsilon}_{\mu_1
\mu_2 \mu_3 \mu_4 } G^{\nu_1\mu_1} G^{\nu_2 \mu_2} G^{\nu_3 \mu_3} G^{\nu_4
\mu_4} G^{55} \equiv  det|g|\phi~{\bf 1},
\end{eqnarray}
where $det|g|$ is the determinant of the 4-dimensional metric and
${\epsilon}$'s are the fully antisymmetric Levi-Civita tensors.
The expression of generalised determinant is rather simple as the metric is
diagonal.
\setcounter{equation}{0}
\section{ Generalized connection, torsion and curvature}
\subsection {Covariant derivative and generalized structure equations}

Following Connes \cite{Co2}, we define the generalized connection through a
covariant derivative $\nabla$. The covariant derivative as a
direct generalization of the ordinary one is an operation which acts on
an 1-form satisfying the properties
\begin{eqnarray}\label{conn}
&& \nabla : \Omega^1\longrightarrow  \Omega^1
\otimes_{{\cal A}} \Omega^1~, \nonumber \\
&& \nabla(U F) = (\nabla U) F + U \otimes D F~.
\end{eqnarray}
Here the tensor product
$\Omega^1 \otimes_{{\cal A}} \Omega^1$ is generated by the
elements $\{ U_1 \otimes U_2 ; U_1, U_2 \in \Omega^1\}$ with the
relation $U_1 F\otimes U_2 = U_1 \otimes F U_2$
for any $F \in \Omega^0$.

Due to the property (\ref{conn}), the covariant derivative of an arbitrary
1-form is known if its action on a basis is given. Hence, the generalized
covariant derivative is equivalently given by a set of generalized connection
one-forms $\Omega^A_{~~B} \in \Omega^1$, the relation being
\begin{equation}\label{conn1}
\nabla E^A = E^B \otimes \Omega^A_{~~B}~.
\end{equation}

The connection is said to be a Levi-Civita or metric compatible connection,
if it satisfies $\nabla G = 0  $. By definition, the covariant
derivative of the metric functional is given by the following rule
\begin{equation}\label{comp}
\nabla G(U,W)= D(<U, W>) + <\nabla U, W> + <U, \nabla W>~,
{}~\forall ~ U, W \in \Omega^1({\cal M})~.
\end{equation}
By imposing the metric compatibility condition in Eq (\ref{comp}) in the
orthonormal
basis (\ref{gviel}) gives
\begin{equation}\label{comp1}
\Omega^{A~~\dagger}_{~~B}~~=~~ - \eta^{AD} \Omega^C_{~~D} \eta_{BC}.
\end{equation}
Explicitly,
\begin{eqnarray}\label{comp2}
\Omega^a_{~b\mu} & = & -\eta^{ad}\Omega^c_{~d\mu}\eta_{cb}, \nonumber \\
\Omega^a_{~b 5} & = & -\eta^{ad}{\tilde \Omega}^c_{~d 5}\eta_{cb}, \nonumber \\
\Omega^a_{~{\dot 5}\mu} & = & -\eta^{ab}\Omega^{\dot 5}_{~b\mu}, \nonumber \\
\Omega^a_{~{\dot 5}5} & = & -\eta^{ab}{\tilde\Omega}^{\dot 5}_{~b5},\nonumber
\\
\Omega^{\dot 5}_{~{\dot 5}\mu} & = & 0, \nonumber \\
\Omega^{\dot 5}_{~{\dot 5}5} & = & f(x)r,
\end{eqnarray}
where f(r) is an ordinary function.

As the connection has been introduced independently of the metric structure,
it cannot be in general determined in terms of the vielbein defined from the
metric.  However, the metric compatibility condition (\ref{comp1}) can be used
effectively as a supplementary condition. We shall employ it along with torsion
free
structure equation to determine the connection.

The generalized Cartan structure equations define torsion and
curvature of a given connection as follows:
\begin{eqnarray}
T^A = D E^A - E^B \wedge \Omega^A_{~~B}~, \label{tors} \\
R^A_{~~B} = D \Omega^A_{~~B} + \Omega^A_{~~C}
\wedge \Omega^C_{~~B}~, \label{curv}
\end{eqnarray}
where $T^A$ and $R^A_{~~B}$ are $2$-forms.

\subsection{ Torsion free connection}
As in the case of the ordinary Riemannian geometry, we can impose the
torsion free condition $T^A = 0$~. Then the structure equation
(\ref{tors}) reduces to
\begin{equation}
D E^A = E^B \wedge \Omega^A_{~~B}~, \label{0tors}
\end{equation}
Eq (\ref{0tors}) and the condition (\ref{comp1}) that follow from metric
compatibility "overconstrain" the connection in our geometry. However, we can
determine it uniquely provided the vielbein (i.e. the metric structure)
satisfies some restrictive conditions. This is in contrast with the Riemannian
geometry, where metric compatibility determines the connection uniquely for any
given metric, if torsion vanishes.

The structure equation (\ref{0tors}) gives us the connection 1-forms
$ \Omega^{\dot 5}_{~~a~}$ and $\Omega^a_{~~b}$, which has the following
components
\begin{eqnarray}\label{CON1T0}
\Omega^{\dot 5}_{~~a~\nu} ~&=&~E^\mu_{~a}( {1\over 2} F_{[\mu \nu]}~ + ~
X_{(\mu \nu)})  \cr
\Omega^{\dot 5}_{~~a~5}~ &=& ~ {\tilde E}^\mu_{~a}(\partial_\mu\Phi~ +~
ma_{-\mu }r ~-~ {\tilde A}_\mu f r)\cr
\Omega^a_{~~b~\nu} ~&=&~E^\rho_{~b}({1\over 2}(\partial_\rho E^a_{~\nu}~-~
\partial_\nu E^a_\rho)-{1\over 2}(A_\rho \Omega^a_{~~5~\nu}~-~A_\nu
\Omega^a_{~~5~\rho})~+~ Y^a_{(\rho \nu )})\cr
\Omega^a_{~~b~5} ~&=&~{\tilde E}^\rho_{~b}( \Phi \Omega^a_{~{\dot 5}~\rho} ~+~
m
e^a_{-\rho}r ~-~{\tilde A}_\rho \Omega^a_{~{\dot 5}~5})
\end{eqnarray}
where
$ F_{[\mu\nu]} ~=~ \partial_\mu A_\nu ~-~\partial_\nu A_\mu $ ~,~
$ u_{\pm} = u_2 \pm u_1 $ for any function $U$. The connection 1-forms
$\Omega^a_{~~{\dot 5}}$ and the symmetric tensor
functions $X_{(\rho\nu)}, Y^a_{~(\rho\nu)}$ are to be determined by the
metric compatible conditions (\ref{comp2}).

{}From Eq (\ref{comp2}) we derive the equations, that determine $X_{(\rho\nu)}$
\begin{eqnarray}\label{XDET}
{\tilde \Phi}{\tilde X}_{(\rho \nu)}~+~\Phi X_{(\rho\nu)} &~=~& -~{1\over 2}
({\tilde F}_{[\rho\nu]}{\tilde \Phi} ~-~ F_{[\rho\nu]}\Phi) ~-~ mr({\tilde
E}^b_{~\rho}e_{-b~\nu} ~-~ E_{a\nu} e^a_{-\rho}
{}~-~  A_{\nu}a_{-\rho}   \nonumber \cr
&~+~& {\tilde A}_\rho a_{-\nu})
{}~+~ {\tilde A}_\rho
\partial_\nu {\tilde \Phi} ~+~ A_\nu \partial_\rho \Phi
\end{eqnarray}
This equation implies two independent equations. One of them is a constraint
on vielbein
\begin{eqnarray}\label{VIELCONST}
e_1 ~& = &~ \beta e_2 ~=~ \beta e ~,\cr
a_1 ~& = &~ \alpha a_2~=~\alpha a ~, \cr
\phi_1~&=&~{\phi_2\over \alpha } ~=~{\phi \over \alpha },
\end{eqnarray}
where $\alpha $ and $\beta$ are two arbitrary functions. In this case,
$e^a_{~\mu}~,~a_\mu~,~\phi~,~\beta$ and $\alpha $ are independent
dynamical variables. ( As we shall see later, the function $f$ in
$\Omega^5_{~~5}$ without having a kinetic term, can be eliminated from the
Lagrangian).
The second equation determines $X_{(\rho \nu )}$
\begin{equation}\label{XXX}
X_{(\rho \nu )}   ~=~x_{(\rho \nu)} \pmatrix{ \alpha & 0 \cr
                                                   0 & 1\cr}~,
\end{equation}
where
\begin{eqnarray}
x_{(\rho \nu)}   &~=~& - {1\over 2\phi} m~(\beta-1)^2 g_{(\rho
\nu)}~+~( a_\nu {\partial_\rho\phi \over \phi} ~+~ a_\rho {\partial_\nu\phi
\over \phi})
{}~-~{m \over 2 \phi} a_\rho a_\nu (\alpha -1 )^2 \nonumber \cr
&~-~&{1\over 4}( a_\nu{\partial_\rho\alpha \over \alpha }~+~a_\rho
{\partial_\nu\alpha \over \alpha}) \nonumber
\end{eqnarray}
With the vielbein satisfying (\ref{VIELCONST}) we can determine the connection
1-form uniquely with the result
\begin{eqnarray}\label{CON2T0}
\Omega^{\dot 5}_{~~{\dot 5}~\mu} &~=~&  0~, \cr
\Omega^{\dot 5}_{~~{\dot 5}~5} &~=~& f.r ~, \cr
\Omega^{\dot 5}_{~~b~\nu} &~=~&{1\over 2} e_a^{~\rho }\Big [ \Big (~f_{[\rho
\nu]}~-~
{m\over \phi}(\beta - 1)^2 g_{\rho \nu}~+~( a_\rho {\partial_\nu\phi\over \phi}
 ~+~ a_\nu {\partial_\rho\phi \over \phi})
{}~-~{m \over \phi}(\alpha-1)^2a_\rho a_\nu \cr
&~-~& {1\over 2}(a_\nu{\partial_\rho \alpha \over \alpha }+
a_\rho {\partial_\nu \alpha \over \alpha} ) \Big )
\pmatrix{\alpha \beta^{-1} & 0 \cr
0 & 1 \cr} + l_{[\rho \nu ]}\pmatrix{\beta^{-1} & 0 \cr
                                     0 & 0 \cr}  \Big ] ~,\cr
\Omega^{\dot 5}_{~~b~5} &~=~& e_b^{~\rho}\Big [ \partial_\rho\phi
\pmatrix{ \alpha ^{-1} & 0 \cr
             0 & \beta^{-1} \cr}
{}~-~ \phi {\partial_\rho \alpha \over \alpha}
\pmatrix{\alpha ^{-1}& 0 \cr
            0 & 0\cr}  \cr
&~-~&a_\rho ~r ~( ~m (\alpha-1)
         \pmatrix{1 & 0 \cr
         0 & \beta^{-1}\cr}~+~ f \pmatrix{1 & 0 \cr
                                  0 & \alpha \beta^{-1}\cr} \Big ] ~, \cr
\Omega^a_{~~b~\mu} &~=~& \omega^a_{~b~\mu}~+~e^{~a\tau}e_b^{~\rho}
\Big [ \Big ( g_{\tau \mu}~ {\partial_\rho \beta \over \beta }~-~g_{\rho \tau}~
{\partial_\mu \beta \over \beta })
     \pmatrix{1 & 0 \cr
                 0 & 0 \cr}
{}~+~\Big (~ a_\mu~ f_{[\rho \tau]} ~+~ a_\mu~ h_{[\tau \rho]}  \cr
&~+~& {1\over 2}l_{[\rho \tau]}
{}~+~ { m \over \phi }(\beta -1)^2 ( a_\tau g_{\mu \rho} -
 a_\rho g_{\mu \tau})\Big )\pmatrix{ \alpha ^2\beta^{-2} & 0 \cr
                                      0 & 1 \cr} \Big ]~, \cr
\Omega^a_{~~b~5} &~=~& {1\over 2\beta} e^{~a\tau} e_b^{~\rho}\Big [
 m~(1-\beta^2)
g_{\rho \tau}~r ~+~
 a_\tau a_\rho~ r ~(~m~(\alpha ^2 -1) ~+~f\alpha)  \cr
&~+~&{3\over 2}~ \phi ~ a_\tau {\partial_\rho\alpha \over
\alpha }~r ~-~{1\over 4}~\phi~ a_\rho {\partial_\tau \alpha \over \alpha }~ r
{}~+~(a_\rho\partial_\tau\phi-a_\tau\partial_\rho\phi)~-~\phi ~f_{[\tau\rho]}
\Big ] ~,
\end{eqnarray}
where
\begin{eqnarray}
\omega^a_{~b \nu} &~=~&{1\over 2}e^{a\tau} ~e_b^{~\rho}\Big [ e^d_{~\tau
}(\partial_\rho e_{d\nu}
{}~-~\partial_\nu e_{d\nu})~+~e^d_{~\rho }(\partial_\nu e_{d\tau} ~-~
\partial_\tau e_{d\nu})~ \cr
&~+~& e^d_{~\nu}(\partial_\rho e_{d\tau}~-~\partial_\tau e_{d\rho })
\Big] ~, \cr
f_{[\tau \rho]} &~=~& \partial_\tau a_\rho ~-~ \partial_\rho a_\tau ~,\cr
h_{[\tau \rho]} &~=~& a_\tau {\partial_\rho \phi \over \phi} ~-~ a_\rho
{\partial_\tau \phi \over \phi}~, \cr
l_{[\tau \rho]} &~=~& a_\tau {\partial_\rho \alpha \over \alpha } ~-~
a_\rho {\partial_\tau \alpha \over \alpha},
\end{eqnarray}
and $\omega^a_{~b}$ is the ordinary metric compatible and torsion free
connection.

\subsection { The generalized Ricci scalar curvature and the action}

 Eq(\ref{curv}) determines completely the curvature 2-forms $R^A_{~~B}$
and hence the Ricci scalar once the set of the connection 1-forms $\Omega^A_{~~
B}$ are given. The Ricci scalar curvature in our case can be defined as follows
\begin{equation}\label{RISCAL}
R ~~= ~~< E^A\wedge E^B ~,~ R_{AB}>
\end{equation}
where $R_{AB} = \eta_{AC}~R^C_{~~B}$. It is convenient to compute the Ricci
scalar curvature in the $DX^M$ basis.
The curvature 2-form $ R_{AB}$ can be expanded in the form
\begin{equation}
R_{AB} ~~= ~~ DX^M\wedge DX^N~ R_{ABMN}
\end{equation}
The $Z_2$ functions $R_{ABPQ}$ are determined uniquely, due to the
anti-commutativity (\ref{WEDGE}) of the diagonal basis $DX^M$. The vielbein
can be also expanded in terms of $DX^M$. Using the sesquilinearity and the
hermicity of the innner product we can bring all the coefficients out and we
are
left with the inner products in $DX^M$ basis as defined in Eq (\ref{INNER3})
It is straightforward to substitute this inner product into Eq (\ref{RISCAL})
and
compute the Ricci scalar curvature.

The action is defined as
\begin{equation}\label{action}
S ~~=~~ {1\over m. \kappa}~Tr~(\int dx^4 \sqrt{-det~G}~ R)~,
\end{equation}
where $\kappa $ is a constant to be fixed later.

The integration over the discrete space follows naturally to be
${1\over m} Tr $.
\setcounter{equation}{0}
\section{ Some particular cases}
It is cumbersome but straightforward to compute the action based on Eq
(\ref{action}). The resulting action involves lengthy expressions
including kinetic terms and
cross-interaction terms for gravity, the vector field $a_\mu(x)$ and the scalar
fields $\beta(x)$, $\alpha(x)$ and $\phi(x)$. The function $f(x)$ in Eq
(\ref{comp2}) appears as an auxilary field without a kinetic term. It can be
eliminated from the action. Rather than present the full action, in what
follows, we shall discuss some special cases that demonstrate the role
$\alpha$ and $\beta$ play and their physical significance.

\subsection{ The zero modes sector of the Kaluza-Klein theory}
In this case we choose $ \beta~=~1~,~\alpha ~=~1$. We recover the action of the
previous paper \cite{LVW}. With the field redefinition
$a \longrightarrow a\phi $, the result is
\begin{equation}
{\cal L}~=~ {1 \over m\kappa}\int d^4x \sqrt{-det|g|} (\phi~r_4 ~-~2 \Box \phi
 ~+~ {\phi^3\over 4} f_{\mu\nu} f^{\mu \nu}),
\end{equation}
where $ r_4$ is the four-dimensional Ricci scalar curvature.
We see that the discretised version contains only the zero mode sector of the
Kaluza-Klein theory.

As $\phi(x)$ and $a_\mu(x)$ are dimensionless we can introduce the dimensional
parameters $v$ and $b$ with dimension of mass into the theory via
\begin{eqnarray}
\phi(x) ~& \longrightarrow &~ e^{{\chi(x) \over v}}  \cr
a_\mu(x)~& \longrightarrow &~  {a_\mu(x) \over b}
\end{eqnarray}
The Lagrangian (5.1) then assumes the form
\begin{equation}
{\cal L}~=~ {1\over m \kappa}\int d^4x \sqrt{-det|g|}e^{{\chi(x)/v}}(r_4
{}~-~{2 \over v^2} \partial_\mu \chi(x) \partial^\mu \chi(x) +
{e^{2\chi(x)\over
v} \over 4b^2} f_{\mu \nu} f^{\mu \nu}),
\end{equation}
and leads to the identification
\begin{eqnarray}
m\kappa~~ &~=~& 16\pi G^2,\cr
4\pi G^2 v^2 &~=~&~~ 1 ,\cr
16\pi G^2 b^2&~=~&~~ 1 ,
\end{eqnarray}
in order to have standard expressions for the kinetic terms.
The scalar field $\chi(x)$ and the vector field $a_\mu(x)$ are massless.

\subsection{Massive dilaton and cosmological constant}

In this subsection, we consider the case, where $ a=0~, \alpha ~=~\phi~=~1$
to see the new features the dynamical variable $\beta(x)$ brings
into the theory.
The Lagrangian in this case reduces to
\begin{eqnarray}
{\cal L} &~=~& {1\over 16\pi G^2}\int d^4x \sqrt{-det|g|}({1\over
2}(\beta^2+1)r_4~+~{3\over 2}\partial_\mu \beta \partial^\mu \beta \cr
&+& {m^2 \over 4 \beta^2}(\beta - 1)^3
(2\beta^6+2\beta^4+7\beta^3-21\beta^2+9\beta-5))
\end{eqnarray}
By redefining $ \beta \longrightarrow {{\tilde \beta }\over u}$, where u is a
parameter with dimension of mass, we obtain a dynamical scalar field ${\tilde
\beta }(x)$.
The Lagrangian now contains gravity and a dilaton with a highly nonlinear
potential
\begin{equation}
{\cal L} = {1\over 16\pi G^2}\int d^4x \sqrt{-det|g|}({{\tilde \beta}^2 +
u^2 \over u^2}r_4 + {3\over 2u^2} \partial_\mu {\tilde \beta } \partial^\mu
{\tilde \beta} + V({ \tilde \beta}))
\end{equation}
To have the right factor for the kinetic term of ${\tilde \beta}$ the
parameter $u$ should be given by
\begin{equation}
u~=~ \sqrt{{3\over m\kappa}}.
\end{equation}
The potential is non-renormalisable as to be expected from a theory of gravity.
The $V({\tilde \beta})$ potential has a minimum at ${\tilde \beta} = u$.
Expanding ${\tilde \beta}$ around $u$, we obtain a mass term for the
${\tilde \beta} $ field.
\begin{equation}
{\cal L}_{mass} ~=~ {95\over 4}m^2{\tilde \beta}^2.
\end{equation}
 Hence, although on one sheet ${\tilde \beta}(x)$ rescales gravity and acts
like a dilaton field of conventional theory, it has a mass in the present
framework.

The mass of the $\beta$ dilaton is $\sim 4.87 m $. If the theory is applied
in the two left- and right-sheeted model of Connes and Lotts \cite{CoLo}
with $m$ as the electroweak scale $\sim 246 GeV$, then the $\beta$ dilaton has
mass in the TeV range.

We can also imagine $\beta(x)$ to be just a constant. In that case $V(\beta)$
plays the role of the cosmological constant that vanishes at $ \beta=1$ ( or
$\tilde \beta = u$). The cosmological constant is positive ( negative) for
$\beta > 1$ ($\beta < 1$). It can be made arbitrarily small by taking its value
arbitrarily close to unity. This feature has clearly important application in
cosmology.

\subsection{ Mass term, quartic potential and higher derivative interactions
for the vector field}

In this subsection we consider the special case, where we have flat space-time,
$\phi(x) =1$. $ \beta $ and $\alpha $ assumed to be constant other
than unity. For simplicity we shall assume $\beta = \alpha $
The Lagrangian in this case is given by
\begin{equation}
{\cal L} ~=~ {1\over 16\pi G^2}\int d^4x \sqrt{-det|g|}
( {\cal L}_{kin}~+~{\cal L}_2~+~{\cal L}_4~+~{\cal L}_{high}~+~const ),
\end{equation}
where
\begin{eqnarray}
{\cal L}_{kin} &~=~& - {1\over 8 \alpha ^2}(2\alpha^3- \alpha^2 -\alpha -2)
f_{\mu\nu} f^{\mu \nu}
, \cr
{\cal L}_2 &~=~& m^2~a_\mu a^\mu {(\alpha -1)^4(\alpha + 1)^3 \over
8 \alpha },\cr
{\cal L}_4 &~=~& m^2 (a_\mu a^\mu)^2
{(\alpha -1)^3(\alpha +1)(\alpha^3+1)(\alpha -{3 \over 2})\over 4\alpha},\cr
{\cal L}_{high}&~=~&{1\over 8} {(\alpha - 1)^2(\alpha + 1) \over \alpha }
(a^\mu f_{\mu\nu})^2.
\end{eqnarray}
The vector field $ a_\mu(x)$ indeed has a nonvanishing mass term.
\setcounter{equation}{0}
\section{ Summary and Conclusions}

The noncommutative geometric approach \`a la Connes has provided new insights
into the Standard Model of elementary particle interactions by providing a
unified geometric desription of gauge and Higgs particles and at the same time
providing a Higgs potential with spontaneously broken symmetry. Since the
approach is based on a fundamentally new structure of space-time, it is natural
to ask how gravity fits into the picture and what are the consequences on the
other interactions. Indeed as shown by one of the authors \cite{MRST} of this
paper, if one assumes the same underlying space-time structure for both
gravity and electroweak interactions, one can predict the top quark and Higgs
particle masses  $ m_t \sim 172 GeV$ and $m_H \sim 241 GeV$.

The present work is an extension of that in \cite{LVW}.  The noncommutative
geometry is based on the algebra ${\cal A}={\cal C}^\infty({\cal M})\times
Z_2$. The discrete $Z_2$ structure supplementing the smooth functions on the
four-dimensional manifold allows one to introduce a Dirac operator that has a
component corresponding to an outer automorphism of the algebra and develop a
differential geometry that has close anology with the usual Riemannian
geometry. The discrete elements belonging to
$Z_2$ may be considered as two discrete points of a fifth spatial dimension in
Kaluza-Klein-type theories or alternately simply as giving rise to two
independent copies of space-time. We have adhered to the first point of view
in this paper.

We have considered in this paper the case of a torsion free,
metric compatible connection. The ensuing lagrangian has a rich and complex
structure with a finite field content including two additional dynamical
scalar fields along with the fields introduced in the vielbein. To understand
the structure in more physical terms, we have considered some special cases.
First of all, if the fields $\alpha $ and $\beta$ are constants equal to unity,
we
obtain the previously studied case \cite{LVW} in which there are only zero mass
tensor, vector and scalar fields. Secondly, if we consider only gravity
and the field $\beta(x)$, we find that while rescaling the metric , $\beta(x)$
acts as a dilaton field of conventional theories, it is different in that it
has a mass term in the present framework. On the other hand if we treat
$\beta(x)$ to be a constant
everywhere, it gives rise to a cosmological constant term that vanishes at
$\beta=1$. It can be positive or negative depending upon whether $\beta>1$
or $\beta<1$ and can be made arbitrarily small by taking $\beta$ arbitrarily
close to unity. Thus, the model can have application in Cosmology. As a third
possibility, if we consider $ \alpha $ and $\beta$ as constants we obtain
a model in which the vector field is massive.

These special cases give rise to a rich variety of physical models. In the
more general case when torsion is present the metric compatibility condition
(\ref{comp1}) is sufficient to determine uniquely a non-vanishing torsion in
terms of the metric without imposing any constraints on the vielbein. ( The
Eq (\ref{VIELCONST}) now instead of being a constraint on the vielbein becomes
a torsion determining equation). Hence, a
pair of independent metric, vector and scalar fields can coexist and we expect
that one field in each pair is massless and the other is massive. We defer the
consideration of this case in a paper to follow \cite{VW}.

Finally, we note that such a general metric with two
independent tensor, vector and scalar fields opens up new and intriguing
possibilities for the Standard Model. In Connes-Lott model of right - and
left-handed sheets, for instance, two different metrics on the two sheets
implies that gravity couples differently to different chiralities. Distance
measured by a left-handed beam is different from that measured by a
right-handed one. Are these considerations relevant to the Standard
Model? We do not know at present. But even the simplest version of NCG based on
two sheets has all these possibilities sugests that it merits further study.
\bigskip

\noindent
{\bf Acknowledgments.}

This work was supported in part by the U.S. Department of Energy under contract
number DE-FG02-85ER40231. G.Landi was involved in the first stage of this
program. We regret that he became more interested in other things than this
very exciting matter.


\bigskip

\begin{thebibliography}{99}
\bibitem{Co1} A.Connes, Publ.Math. IHES {\bf 62} (1983) 44.
\bibitem{Co2} A.Connes, Publ.Math. IHES M/93/12 (1993) (to be published by
Academic Press).
\bibitem{CoLo} A.Connes and J.Lott, Nucl.Phys. {\bf B18} (Proc.Suppl.) (1990)
29.
\bibitem{ChamFF1} A.H.Chamseddine, G.Felder and Fr\"ohlich, Nucl.Phys.{\bf
B395} (1993) 672.
\bibitem{BGW} B.Balakrishna, F.G\"ursey and K.C.Wali, Phys.Rev. {\bf D44}
(1991) 3313.
\bibitem{KK} Th.Kaluza, Sitzungsber. Preuss. Akad. Wiss. Phys.
Math. Klasse 966 (1921); \\
O.Klein, Z.F. Physik {\bf 37} (1926) 895; \\
Y.Thirry, Comptes Rendus (Paris) {\bf 226} (1948) 216.
\bibitem{ChamFF2} A.H.Chamseddine, G.Felder and Fr\"ohlich, Comm.Math.Phys.{\bf
155} (1993) 201.
\bibitem{Kast} D.Kastler, {\it The Dirac Operator and Gravitation} Marseille
Preprint CPT-93/p.2970 (1993).
\bibitem{KaWa} W.Kalau and M.Walze, {\it Gravity, Non-Commutative Geometry and
the Wodzicki Residue} Preprint MZ-TH/93-38 (1993).
\bibitem{SI} A.Sitarz, {\it Gravity From Noncommutative Geometry} Preprint
TPJU-1/1994, hep-th/9401145 (1994).
\bibitem{China} B.Chen, T.Saito and K.Wu, {\it Gravity and Discrete Symmetry}
ASITP Preprint, April (1994).
\bibitem{LVW} G.Landi, Nguyen Ai Viet and K.C.Wali, Phys.Lett. {\bf B326}, 45.
(1994).
\bibitem{Klim}C.Klim\~c\'ik, A.Pompo\~s and V.Sou\~cek, Lett.Math.Phys. {\bf
30}, (1994), 259.
\bibitem{Madore} M.Dubois-Violette, R.Kerner and J.Madore, Class.Quantum Grav.
{\bf 6} (1989), 1709; J.Math.Phys. {\bf 31}(2) (1990), 316 ; J.Madore,
Phys.Rev. {\bf D41} (1990), 3709.
\bibitem{PoDu} M.J.Duff, C.N.Pope and K.S.Stelle, Phys.Lett.{\bf B223} (1989),
71.
\bibitem{KUMA} G.Kunstatter, J.W.Moffat and J.Malzan, J.Math.Phys.{\bf 24}
(1983), 886; R.B.Mann, Nucl.Phys.{\bf B231} (1983), 481.
\bibitem{COQUEV} R.Coquereaux, G.Esposito-Far\`ese and G.Vaillant, Nucl.Phys.
{\bf B353} (1991), 689.
\bibitem{MRST} Nguyen Ai Viet {\it  Predictions of Non-commutative space-time},
Talk given at MRST meeting, McGill University, Montr\'eal, Canada, 1994 ( To
be appeared in the Proceeding of MRST'94, World Scientific Publishing House,
Singapore, 1995)
\bibitem{VW} Nguyen Ai Viet and K.C.Wali (in preparation).
\end{thebibliography}
\end{document}